\documentclass[aps,prl,twocolumn,showpacs,superscriptaddress,groupedaddress]{revtex4-1}

\usepackage{graphicx}  
\usepackage{dcolumn}   
\usepackage{bm}        
\usepackage{amssymb}   
\usepackage{epstopdf}

\newcommand{\first}{$1^{\mathrm{st}}~$}
\newcommand{\second}{$2^{\mathrm{nd}}~$}

\begin{document}
\title{Ultrafast electron diffraction using an ultracold source}

\author{M.W. van Mourik}
\author{W.J. Engelen}
\author{E.J.D. Vredenbregt}
\author{O.J. Luiten}
\affiliation{Department of Applied Physics, Eindhoven University of Technology, P.O. Box 513, 5600 MB Eindhoven, The Netherlands}

\begin{abstract}
We present diffraction patterns from micron-sized areas of mono-crystalline graphite obtained with an ultracold and ultrafast electron source. We show that high spatial coherence is manifest in the visibility of the patterns even for picosecond bunches of appreciable charge, enabled by the extremely low source temperature ($\sim$ 10 K). For a larger, $\sim$ 100 $\mathrm{\mu m}$ spot size on the sample, spatial coherence lengths $>$ 10 nm result, sufficient to resolve diffraction patterns of complex protein crystals. This makes the source ideal for ultrafast electron diffraction of complex macromolecular structures such as membrane proteins, in a regime unattainable by conventional photocathode sources. By further reducing the source size, sub-$\mathrm{\mu m}$ spot sizes on the sample become possible with spatial coherence lengths exceeding 1 nm,  enabling ultrafast nano-diffraction for material science.
\end{abstract}

\maketitle

The fast pace at which the new field of ultrafast structural dynamics is currently evolving is largely due to spectacular developments in ultrafast X-ray \cite{Emma_NP_10,Chapman_N_11,Boutet_S_12} and electron \cite{Sciaini_RPP_11,Chergui_C_09,Gao_N_13} beams. A particularly interesting development is the ultracold electron source, which is based on near-threshold photo-ionization of a laser-cooled and trapped atomic gas \cite{Claessens_PRL_05,Claessens_PP_07,Taban_EPL_10,McCulloch_NP_11,Engelen_NC_13,McCulloch_NC_13}. Recently it was shown that the ultracold electron source can be operated at femtosecond timescales while, surprisingly, retaining its high spatial coherence \cite{Engelen_NC_13,McCulloch_NC_13}.
Here we present the first diffraction patterns produced by electron bunches generated by such an ultracold and ultrafast source. Even when focusing the electron beam to a micron-sized spot on a graphite sample, we maintain high-visibility diffraction patterns. This is not possible with femtosecond beams generated with conventional planar photocathodes, which lack the required coherence.
This opens the door to new possibilities, such as few-shot femtosecond electron diffraction of membrane protein crystals and ultrafast nanodiffraction, with the prospect of real-time monitoring of biomolecular dynamics.

Typical ultrafast electron diffraction (UED) experiments are performed using a planar photocathode source \cite{Sciaini_RPP_11}, characterized by effective electron temperatures $T\geq 1000$ K. Kirchner \textit{et al.} \cite{Kirchner_NJP_13} have shown that by focusing the femtosecond photoemission laser to a small spot, the root-mean-square (rms) source size can be reduced to $\sigma_{\mathrm{source}} =3~\mathrm{\mu m}$. By combining this with an rms beam size at the sample of $\sigma_{\mathrm{sample}}=77~\mathrm{\mu m}$, they achieved sizeable coherence lengths,
\begin{equation}
\label{eq:coherence_source}
L_\perp=\frac{\hbar}{\sqrt{m k_\mathrm{B} T}}\frac{\sigma_{\mathrm{sample}}}{\sigma_{\mathrm{source}}}\approx 20~\mathrm{nm},
\end{equation}
with $\hbar$ Dirac's constant, $m$ the electron mass and $k_\mathrm{B}$ Boltzmann's constant. To resolve a diffraction pattern, $L_\perp$ should be larger than the lattice spacing $a$ of the sample under investigation. Therefore, $L_\perp=~$ 20 nm is more than sufficient for protein crystal diffraction (typically $a= 1-5~$nm), as shown in Ref. \cite{Kirchner_NJP_13} on an organic salt with $a\approx~$1 nm.
Unfortunately, the crystallites in thin-film samples for transmission experiments are often limited to (sub-)micron sizes due to limitations in sample synthesis. Conventional photocathode sources cannot attain the required coherence length for $\sigma_{\mathrm{sample}}\lesssim 1~\mathrm{\mu m}$.

The ultracold and ultrafast electron source, previously shown to have source temperatures as low as 10 K \cite{Engelen_NC_13,McCulloch_NC_13}, provides a means to realize ultrafast diffraction on (sub-)micron sized samples.

\begin{figure*}[t]
\centering
\includegraphics[width=\linewidth]{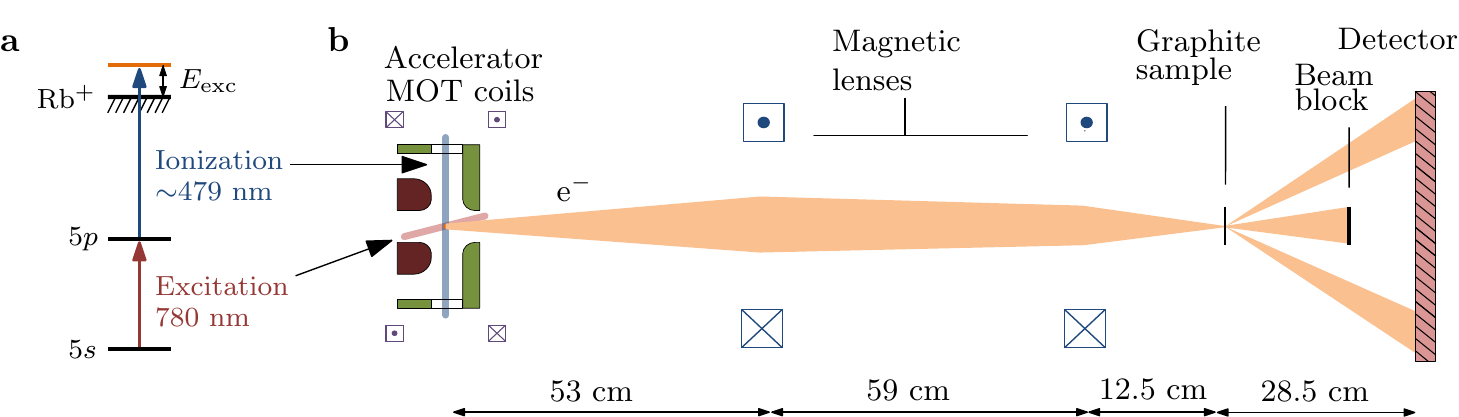}
\caption{Experimental set-up. a). Magneto-optically trapped atoms are ionized by two perpendicular laser beams via a two step ionization scheme. Electrons are first excited from the 5$s$ to the 5$p$ state, and subsequently ionized with a laser pulse. b) The electron bunch is accelerated towards a detector. A set of magnetic lenses controls the beam divergence and waist position. The beam passes through a graphite sample and undergoes diffraction. The 0$^{\mathrm{th}}$ order beam is blocked.}
\label{fig:exp-set-up}
\end{figure*}

Fig.~\ref{fig:exp-set-up} shows a schematic overview of the set-up, which is described in more detail in Refs. \cite{Taban_EPL_10,Engelen_U_14}. Electrons are created by near-threshold photoionization of a laser-cooled and trapped cloud of $^{85}$Rb atoms. Rubidium atoms are first excited from the 5s to the 5p state and subsequently ionized by a $\leq 100$ fs full-width-at-half-maximum (FWHM) long laser pulse with a tunable central wavelength $\lambda_l$ (Fig \ref{fig:exp-set-up}a). Typically, a few hundred electrons are produced per shot. The laser-cooled gas cloud is trapped inside an accelerator structure (Fig. \ref{fig:exp-set-up}b). Electrons are extracted from the cloud by an electric field with strength $F$, and are accelerated to a final energy $U=eFd_{\mathrm{acc}}$, with $e$ the elementary charge, and $d_\mathrm{acc}=12.7~\mathrm{mm}$.
The combination of $\lambda_l$ and $F$ determines the kinetic energy distribution of the released electrons, and thus the effective source temperature $T$ \cite{Engelen_U_14}.
Using the waist scan method (see Methods and \cite{Engelen_NC_13}) we have established that the source temperature $T$ can be varied from 300 K to 10 K by tuning $\lambda_l$ from 477 nm to 500 nm at $F=~$0.85 MV/m.
The source size $\sigma_{\mathrm{source},x(y)}=32\pm 2\:(54\pm 2)~\mathrm{\mu m}$ has been measured by means of an ion space charge scan (see Methods and \cite{Engelen_U_14}). The combination of $\sigma_{\mathrm{source}}$ and $T$ fully characterizes the source.

At 1.245 m from the source, the beam is sent through a 13 -- 20 nm thick monocrystalline graphite sample \cite{Naturally_Graphite} on a 200 mesh copper TEM grid. The detector is placed at a distance $h=0.285$~m from the sample.
Two magnetic lenses (at 0.53 m and 1.12 m) provide control over the spot size and the angular spread of the beam on the sample. To obtain sharp diffraction patterns, we focus the beam on the detector, resulting in a converging beam going through a relatively large area on the sample. Alternatively, focusing the beam on the sample, as shown in Fig. \ref{fig:exp-set-up}b, allows us to investigate the smallest spot size that can be used in ultrafast diffraction. 

Our diffraction images are the result of $10^3$ shots acquired at a 100 Hz repetition rate. Each shot contains a few hundred electrons. For illustrative purposes the recorded diffraction patterns shown here are an average over 10 images. Analyses of data are, however, based on separate $10^3$-shot images.

\begin{figure}[t]
\centering
\includegraphics[width=\linewidth]{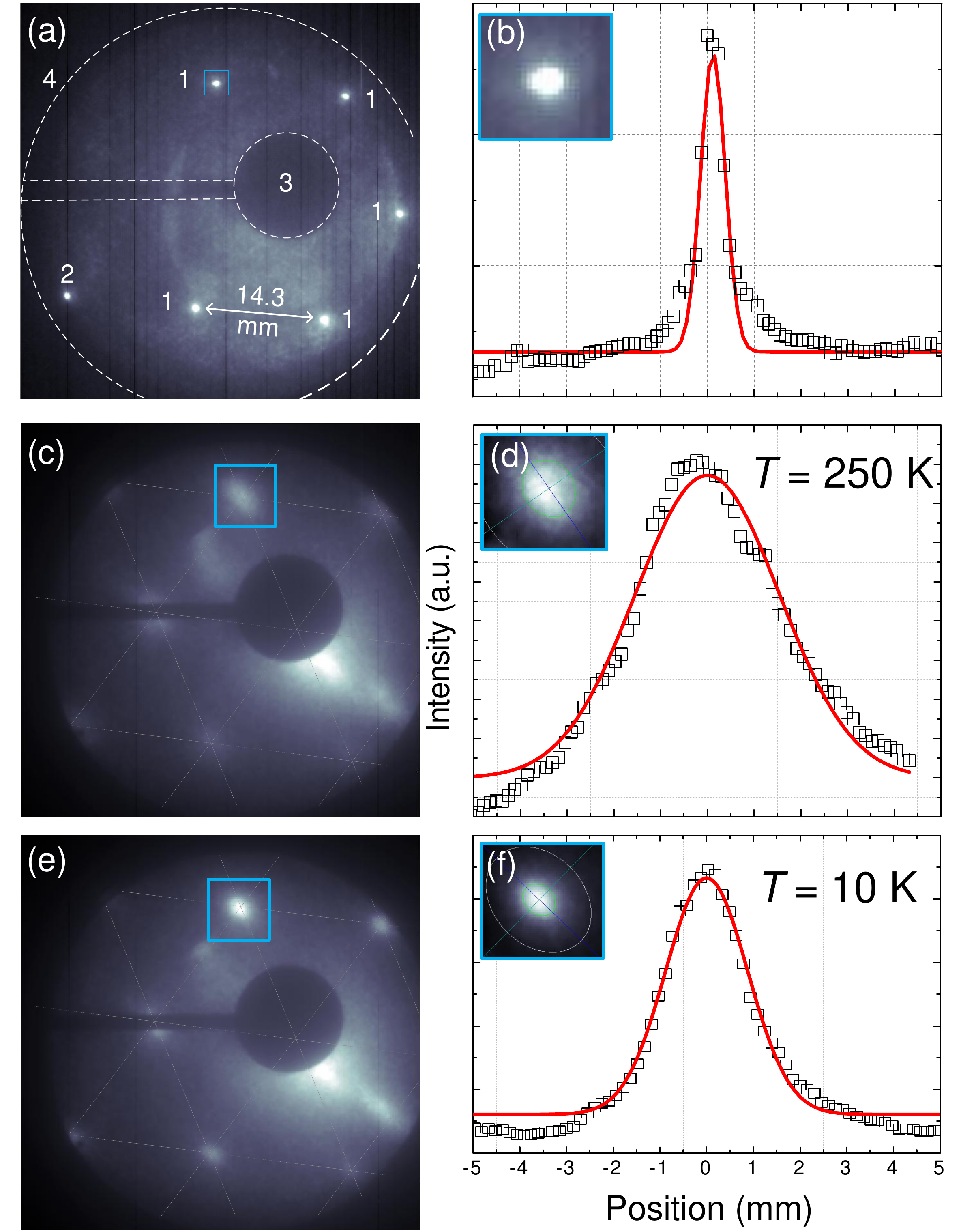}
\caption{Diffraction images obtained with the ultracold source (a) 13.2 keV electrons are focused on the detector using the first magnetic lens. Five \first order spots (1) and one \second order spot (2) are visible. The intraspot distance is 14.3 mm. The beam block (3) and detector edge (4) are outlined. (c) and (e) show diffraction images obtained with a 10.8 keV beam focused on the sample, for source temperatures of $T =~ $250 K (c) and $T =~ $10 K (e). The improved beam quality due to lower temperatures can be seen by comparing close-up views of a spot and the respective line profile of their short axes (d) and (f).}
\label{fig:diffr_images}
\end{figure}

Fig. \ref{fig:diffr_images}a shows an electron diffraction pattern produced with the waist of the beam on the detector, using only the first lens. The pattern was recorded with beam parameters $U=13.2$~keV and $\lambda_l=485$~nm.

Five of the six \first order spots (1) of the expected hexagonal pattern are visible, centered around the beam block (3). The sixth \first order spot is blocked by the stem of the beam block. In the bottom left, a \second order spot (2) can be seen; the others fall outside the detection area (4). The \first order beamlets arrive at the detector at a distance of $s = 14.3~$mm from the central (0$^{\mathrm{th}}$ order) beam. The \first order diffraction angle is $\theta = \tan^{-1}(s/h) = 50\pm 1~\mathrm{mrad}$, in excellent agreement with the theoretical value from Bragg's law,  $\theta= \sin^{-1}(\lambda_e/a_1)=49.9~$mrad. Here $\lambda_e=2\pi\hbar/\sqrt{2m U}$ is the electron's De Broglie wavelength, and $a_1=0.2131~\mathrm{nm}$ is the first order lattice constant of graphite.

The rms spot size on the detector (magnified and profiled in Fig. \ref{fig:diffr_images}b) is measured to be $\sigma_{\mathrm{d},x(y)}=180~(210)~\mathrm{\mu m}$, with x (y) the minor (major) axis of the elliptical spot. The diffraction spots can be brought 12.9 times closer to each other before spot visibility decreases to 88\%. This implies that we could resolve diffraction patterns with lattice distances 12.9 times larger than that of graphite, such as (macro)molecular crystals, with $a$ up to 2.7 nm. The size $\sigma_{\mathrm{d}}$ of the diffraction spot is actually expected to be as small as 30 $\mathrm{\mu m}$, on the basis of measured source temperature and size, but is limited by the detector resolution. Had this not been an issue, crystals with even larger lattice spacings could be studied, up to $a=19$~nm.

We wish to unambiguously demonstrate the full quality of the beam without being limited by detector resolution.
Therefore, we have done measurements with the beam focused to micron-sized spots on the sample (Fig.~\ref{fig:exp-set-up}b). In this configuration diffraction spots expand to a much larger size, but remain distinguishable; a direct consequence of the low-temperature properties of the source. 
For an electron energy $U=10.8$~keV, diffraction images have been taken for ionization laser wavelengths  $\lambda_l=500-476$~nm ($T=10-300$). From GPT particle tracking simulations \cite{GPT}, we find a spot size on the sample $\sigma_{\mathrm{sample}}=3.3~\mathrm{\mu m}$ for 10 K and $\sigma_{\mathrm{sample}}=8.9~\mathrm{\mu m}$ for 300 K.

Figs. \ref{fig:diffr_images}c and e show two examples of diffraction images from this data set, at ionization laser wavelengths of 478 and 498 nm, respectively, corresponding to measured source temperatures of $T$ = 250 and 10 K. The thin gray lines are guides to show the hexagonal diffraction pattern. The spots inside the blue squares have been magnified and profiled in (d) ($T=250$ K) and (f) ($T=10$ K). Two-dimensional Gaussian fits are used to determine the size of the spots. For (d) and (f) these are $\sigma_{\mathrm{d},x(y)}= 1.8~(1.6)~\mathrm{mm}$ and $\sigma_{\mathrm{d},x(y)}=1.1~(0.88)~\mathrm{mm}$, respectively.

\begin{figure}[t]
	\centering
	\includegraphics[width=\linewidth]{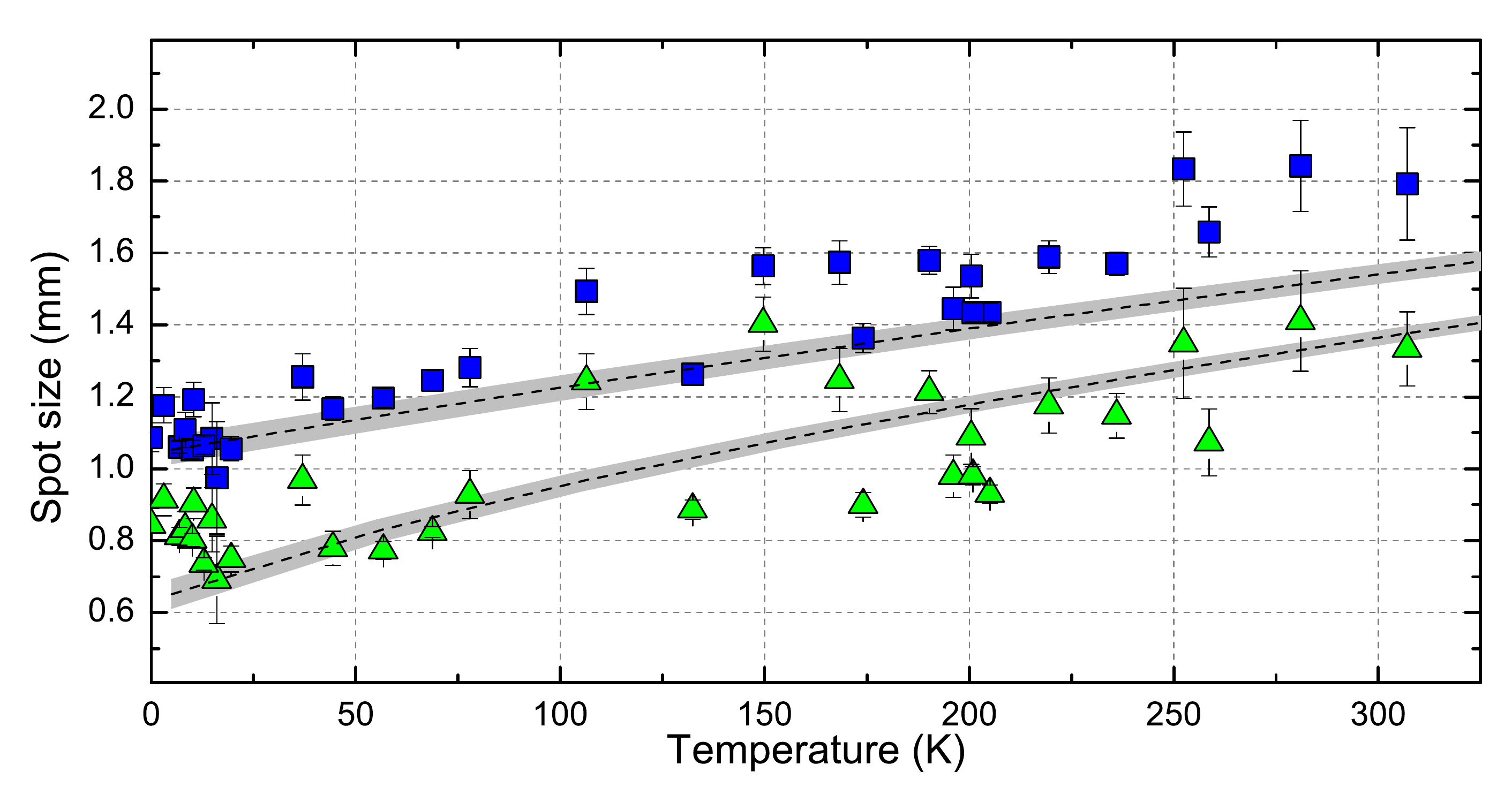}
	\caption{Final diffraction spot size $\sigma_\mathrm{d}$ as a function of effective source temperatures. The two data sets (green triangles and blue squares) are diffraction spot sizes determined using a 2-dimensional Gaussian fit of the elliptical diffraction spots. The gray bands are simulated spot sizes.}
	\label{fig:spotsize}
\end{figure}

\begin{figure}[h]
\centering
\includegraphics[width=\linewidth]{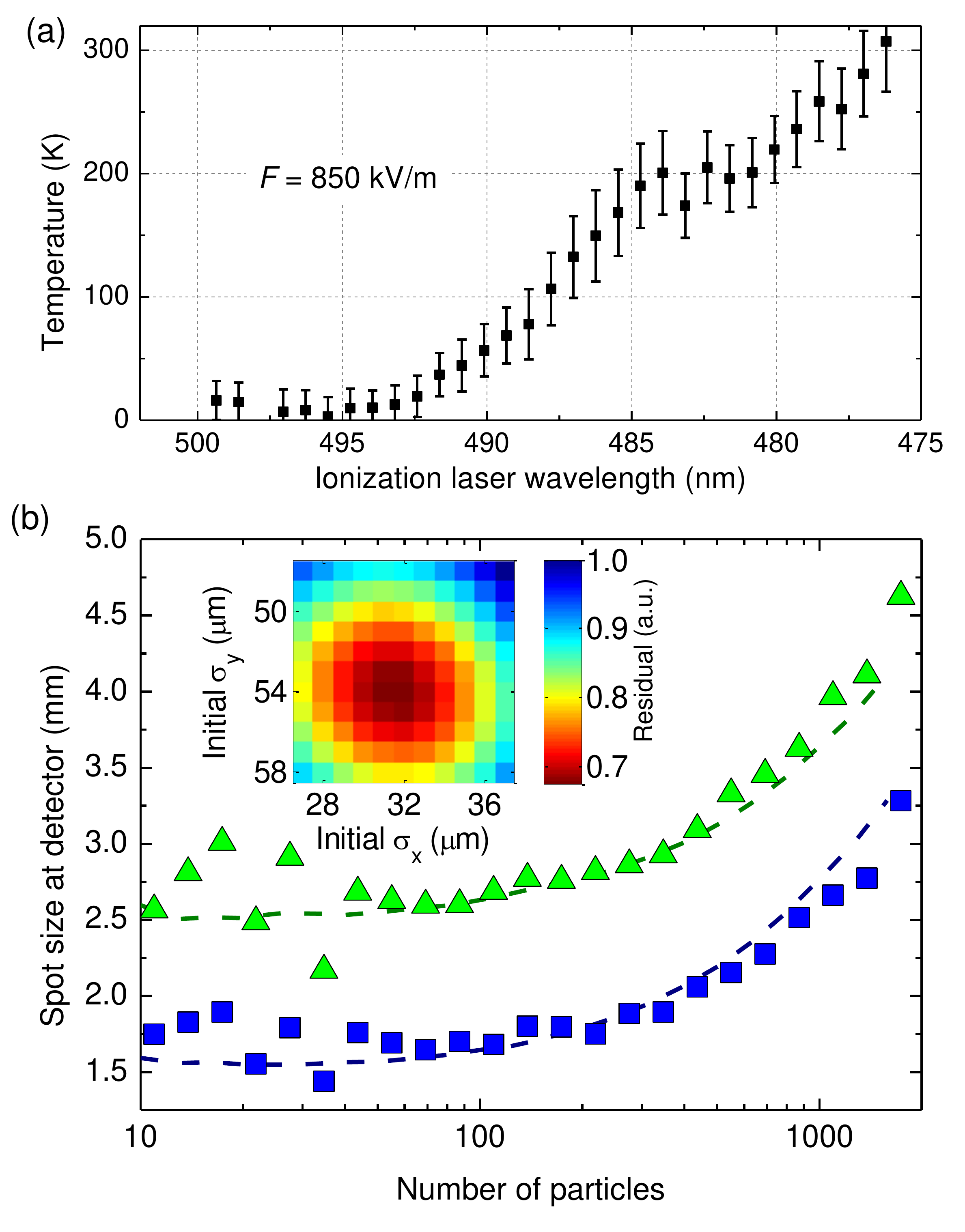}
\caption{Source parameters. (a) Effective transverse source temperature as function of ionization laser wavelength, for an electric field strength of $F=$ 850 kV/m. For large wavelengths, temperatures reach $T \approx$ 10 K. The uncertainty of the data points is partly due to a systematic error in fitting a waist scan. (b) Results of an ion space charge scan to determine $\sigma_{\mathrm{source}}$. The spot size at the detector, in two dimensions (green triangles and blue squares), is shown as function of bunch charge. The inset shows the normalized residual between experimental and simulation data for various simulated source sizes. From this we determine the source size, of which simulated final spot sizes are also plotted (dotted lines).}
\label{fig:waistscan_and_source_size}
\end{figure}

The diffraction spot sizes $\sigma_{\mathrm{d}}$ are plotted as function of source temperature in Fig. \ref{fig:spotsize}, where the two sets represent the rms sizes of the short (green triangles) and long (blue squares) axes of the elliptical spots. Each individual data point is the average over spot sizes obtained from 10 diffraction images. The results are in good agreement with the values from particle tracking simulations (dotted line $\pm~$shaded area), which are based on the measured source temperature and spot size shown in Fig. \ref{fig:waistscan_and_source_size} and the known electric and magnetic fields in the beam line \cite{Engelen_NC_13}. This shows that the spot sizes of the diffraction patterns behave as expected on the basis of source properties. The scatter in the data points is attributed to pointing instabilities in the femtosecond ionization laser, which causes the position and size of the ionization volume, thus the final spot size, to vary.

It is instructive to discuss the results shown in Fig. \ref{fig:spotsize} in terms of coherence length: since $\sigma_{\mathrm{sample}}\ll\sigma_\mathrm{d}$, $\sigma_\mathrm{d}$ is dominated by the angular spread of the beam. This allows us to write $L_\perp$ as
\begin{equation}
\label{eq:L_from_spot}
L_\perp =\frac{a_1 s}{2\pi\sigma_\mathrm{d}},
\end{equation}
implying that $L_\perp$ can be determined directly from diffraction data, independent of the source parameters. For the 10 K data we thus find for the short (long) axis of the elliptical spot $L_{\perp}=~$0.65 (0.48) nm, which drops by a factor 1.5 for the 300 K data. GPT simulations show that $\sigma_{\mathrm{sample}}$ increases by a factor 2.7 for an increase in $T$ from 10 K to 300 K. The values of $L_\perp$ calculated using Eq. (\ref{eq:L_from_spot}) are therefore consistent with Eq. (\ref{eq:coherence_source}), i.e. $\sqrt{T}\propto\sigma_{\mathrm{sample}}/L_\perp$. 

In conclusion, we have produced sharp diffraction patterns using an ultracold and ultrafast electron source. From the quality of these patterns we infer that our ultracold source is suitable for ultrafast diffraction of macromolecular crystals with large ($>1~\mathrm{nm}$) lattice constants. Furthermore, we have shown that we retain high visibility diffraction patterns even if we focus the beam down to a 3 $\mathrm{\mu m}$ spot size on the sample. By varying the source temperature we have shown that our results are consistent with theoretical and simulated models. We can thus extrapolate that for similar beam source and sample sizes, a high temperature ($T\geq 1000~$K) source would have a coherence length around 50 pm, no longer sufficient for diffraction of even the most basic crystals. We have thus shown the advantage of using an ultracold source over conventional photocathodes.

We have used a source size of a few tens of microns across, but this can, at least in principle, be reduced to the same 3 $\mathrm{\mu m}$ used by Kirchner \textit{et al.} \cite{Kirchner_NJP_13}. Combined with a source temperature of 10 K, this would enable us to focus the beam down to $\sigma_{\mathrm{sample}}\approx 300$ nm, and to study sub-$\mathrm{\mu m}$-sized samples, while still maintaining the visibility of Fig. \ref{fig:diffr_images}e. In addition, for a 3 micron source size, time-resolved diffraction of micron-sized protein crystals becomes possible.
\newline
\linebreak
\textbf{Methods}\newline
\textbf{Ionization process}
Rubidium atoms are ionized in a two-step process: Rb atoms are excited from the 5s to the 5p state by an excitation laser pulse, and are subsequently ionized by a laser pulse with tunable ionization wavelength $\lambda_l$.
The excitation and ionization laser pulses propagate along perpendicular directions and overlap in a well-defined region within the magneto-optical trap (MOT), resulting in an ionized cloud with a volume governed by the two laser beam sizes. 
In the photo-ionization process, a few hundred electrons are released with an excess energy given by
\begin{equation}
E_{\mathrm{exc}}=2\pi\hbar c\left(\frac{1}{\lambda_{l}}-\frac{1}{\lambda_0}\right)+2E_h\sqrt{\frac{F}{F_0}},
\end{equation}
with $c$ the speed of light, $\lambda_0=~$479.06 nm, the zero-field ionization threshold wavelength, $E_h=~$27.2 eV the Hartree energy, $F$ the electric field strength inside the accelerator, and $F_0=5.14\times 10^{11}$ V/m the atomic unit of field strength.
The excess energy $E_\mathrm{exc}$ of an absorbed photon is mostly transferred to kinetic energy of the released electron. Due to its broadbanded nature, a femtosecond pulse still has a possibility of ionizing even if the mean photon energy is below the required ionization energy.

\textbf{Source temperature - }In the paraxial approximation, electron trajectories at any point along the beam line can be modelled by a 2~$\times$~2 transfer matrix $\textbf{M}$, which is a known function of magnetic and electric field strengths and positions of various lens elements present in the beam line. The final spot size $\sigma_{\mathrm{d}}$ can be described in terms of source parameters as:
\begin{equation}
\label{eq:raymodel}
\sigma_{\mathrm{d}}^2=M_{11}^2 \sigma_{\mathrm{source}}^2+M_{12}^2\sigma_{\theta,\mathrm{source}}^2
\end{equation}
where $\sigma_{\mathrm{source}}$ and $\sigma_{\theta,\mathrm{source}}$ are the root-mean-square (rms) source size and angular spread, respectively. The angular spread at the source is related to the electron temperature according to $\sigma_{\theta,\mathrm{source}}=\sqrt{k_\mathrm{B}T/2U}$. The values of $M_{11}$, $M_{12}$ are determined via ray tracing models, and $\sigma_{\mathrm{d}}$ is measured at the detector. In the so-called waist scan method, $\sigma_\mathrm{d}$ is measured as function of the focusing strength of a magnetic lens, thus changing the values of $M_{11}$, $M_{12}$, and $\sigma_{\mathrm{d}}$. The resulting data are fitted to Eq. \ref{eq:raymodel}, yielding $\sigma_{\theta,\mathrm{source}}$ and thus $T$. Fig. \ref{fig:waistscan_and_source_size}a shows temperatures determined using waist scans, for excitation laser wavelength $\lambda_l=500-476~\mathrm{nm}$ and electric field strength $F=0.85~\mathrm{MV/m}$.

\textbf{Source size - }The source size is determined by means of an ion space charge scan, in which the spot size of an ion bunch is measured at the detector as function of bunch charge. The spot size is partly governed by the repulsive effects of space charge. Ions are used instead of electrons primarily because the former are negligibly heated during the ionization process, so that angular spread due to temperature can be ignored. We vary the bunch charge by changing the intensity of the ionization laser pulse using neutral density (ND) filters. Fig. \ref{fig:waistscan_and_source_size}b shows the result of a space charge scan. The resulting spot sizes (green triangles and blue squares) are compared to data found with GPT simulations \cite{GPT} (dotted lines). Here, the simulation data has as variable parameters the initial source size in two dimensions, $\sigma_{\mathrm{source,\{x,y\}}}$, and a proportionality factor between laser intensity and bunch charge. The best overlap in experimental and simulation data is found in the least-squares sense, seen in the inset, providing the initial source size, $\sigma_{\mathrm{source},x(y)} = 32\pm 2~(54\pm 2)~\mathrm{\mu m}$.

\noindent\textbf{Acknowledgements}
\newline
This research is supported by the Dutch Technology Foundation STW, applied science division of NWO and the Technology Programme of the Ministry of Economic Affairs.
\linebreak
\noindent\textbf{Author contributions}
M.W.v.M. and W.J.E. performed the experiments and analysed the data; M.W.v.M. wrote the manuscript, with help from E.J.D.V. and O.J.L.; the project was conceived and supervised by E.J.D.V. and O.J.L.

\end{document}